# Coherence properties of a single dipole emitter in diamond

Graham D. Marshall,[1,*] Torsten Gaebel,[2] Jonathan C. F. Matthews,[3] Jörg Enderlein,[4] Jeremy L. O'Brien,[3] James R. Rabeau[2,†]

[1]*Centre for Ultrahigh-bandwidth Devices for Optical Systems, MQ Photonics Research Centre, Centre for Quantum Science and Technology, Department of Physics and Astronomy Macquarie University, NSW 2109, Australia.*
[2]*MQ Photonics Research Centre, Centre for Quantum Science and Technology, Department of Physics and Astronomy, Macquarie University, NSW 2109, Australia.*
[3]*Centre for Quantum Photonics, H. H. Wills Physics Laboratory & Department of Electrical and Electronic Engineering, University of Bristol, Merchant Venturers Building, Woodland Road, Bristol, BS8 1UB, UK.*
[4]*Georg August University Department of Physics III. Institute of Physics Friedrich-Hund-Platz 1 37077 Goettingen Germany.*

On-demand, high repetition rate sources of indistinguishable, polarised single photons are the key component for future photonic quantum technologies [1]. Colour centres in diamond offer a promising solution, and the narrow line-width of the recently identified nickel-based NE8 centre makes it particularly appealing for realising the transform-limited sources necessary for quantum interference. Here we report the characterisation of dipole orientation and coherence properties of a single NE8 colour centre in a diamond nanocrystal at room-temperature. We observe a single photon coherence time of 0.21 ps and an emission lifetime of 1.5 ns. Combined with an emission wavelength that is ideally suited for applications in existing quantum optical systems, these results show that the NE8 is a far more promising source than the more commonly studied nitrogen-vacancy centre and point the way to the realisation of a practical diamond colour centre-based single photon source.

Unwanted photon number states in the spontaneous parametric downconversion (SPDC) process are a detrimental source of noise in current multiphoton experiments [2]. Together with its non-deterministic nature, reliance on SPDC as a source of photons is a major limiting factor in realising scalable photonic quantum technologies. Leading alternatives include colour centres in solids [3] and quantum dots [4-8]. At cryogenic temperatures, quantum dots fabricated in micro-pillar cavities are capable of emitting transform limited single photons to demonstrate quantum interference in the two photon Hong-Ou-Mandel effect [9]. The nitrogen-vacancy centre in diamond (NV) has gained attention in quantum technologies and biological imaging [10-13] for its high quantum yield at room temperature and the polarisation sensitive optical transition of NV that enables manipulation and readout of single electron and nuclear spins in diamond. In the context of *optical* quantum science however, NV has a number of limitations. An isolated centre typically produces single photons in a broad emission spectrum ~100 nm wide with relatively low coupling to the zero-phonon line (ZPL) (the Debye-Waller factor is 0.04 [14])—dispersion makes such a spectrally broad source of light unsuitable for fibre transmission and as with all free-space single emitters, the collection efficiency is poor. The room temperature NV centre excited state lifetimes (typically 12-25 ns [15]) are large in comparison to the femtosecond coherence time for the whole wavelength emission band and even the picosecond coherence time for the ZPL alone [16,17], making it extremely hard to realise a Fourier transform-limited photon source (*i.e.* a source of indistinguishable photons) based on NV.

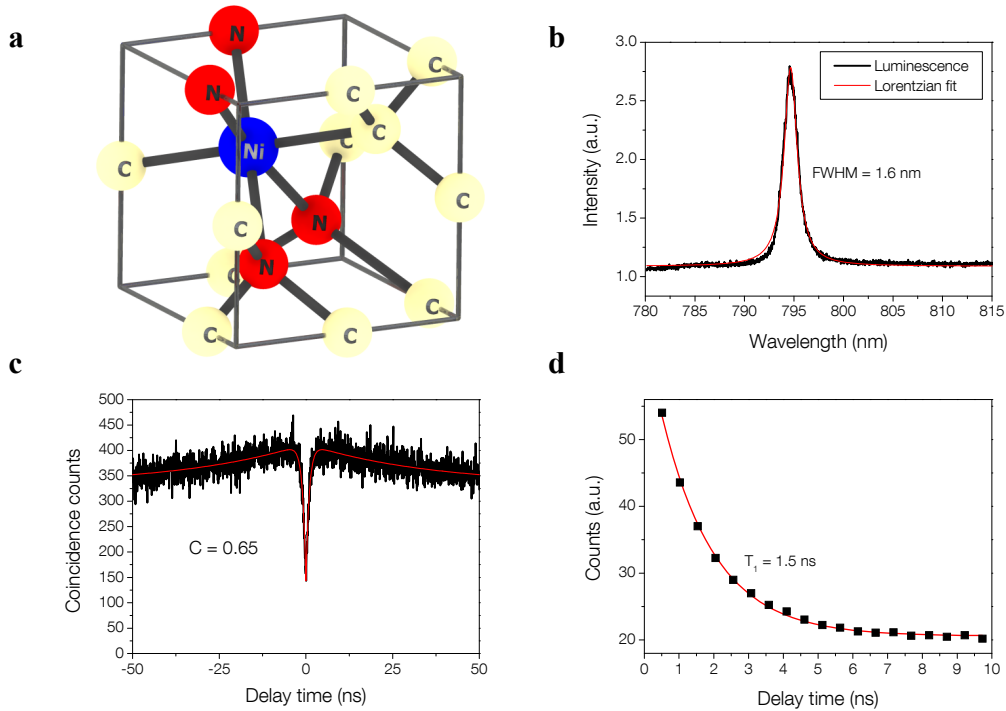

**Figure 1 | The NE8 Colour centre. a**, The structure of NE8. **b**, The luminescence spectrum of NE8 (black curve) with a Lorentzian fit (red curve). **c**, Second order correlation function of the luminescence signal indicating the presence of a single photon emitter. **d**, The luminescence emission as a function of time observed using a pulsed excitation source.

Interest has centred around *single* nickel (Ni) related defects first detected in natural diamond [14] and subsequently synthesised in the laboratory using chemical vapour deposition [18]. Further work has generated a rich area of study, with numerous colour centres at a range of spectral positions both with Ni [19,20] and other defect sites [21] as well as low-temperature investigations of near IR colour centres [22]. The so-called NE8 centre (Figure 1a) is best documented in ensembles [23,24] and *ab initio* density functional theory suggests it is the most stable experimentally measured form of Ni-N complex [25]. NE8 typically has a narrow spectral emission with a full-width half-maximum of less than 3 nm at room temperature and a centre wavelength of approximately 795 nm. These properties make it ideally suited as a replacement for existing SPDC systems that generate photons around 800 nm and where commercially available silicon single photon avalanche photo diodes have high sensitivity. With its demonstration as a room temperature triggered single photon source [26] NE8 is a promising candidate to advance optical quantum technologies. However in order to determine the feasibility of engineering an NE8 based transform-limited source it is essential to be able to measure the orientation of the dipole emitter and characterize the coherence and excited state lifetime properties of the NE8 centre.

We report the results obtained from a single photostable NE8 centre in chemical vapour deposition (CVD) grown diamond nanocrystal and measured using the scanning confocal microscope setup shown in Figure 2a (experimental details are included in Methods). The measured NE8 luminescence spectrum (Figure 1b) consisted of a single line centred at 794.7 nm which is consistent with the expected emission wavelength from the NE8 colour centre in diamond [27].

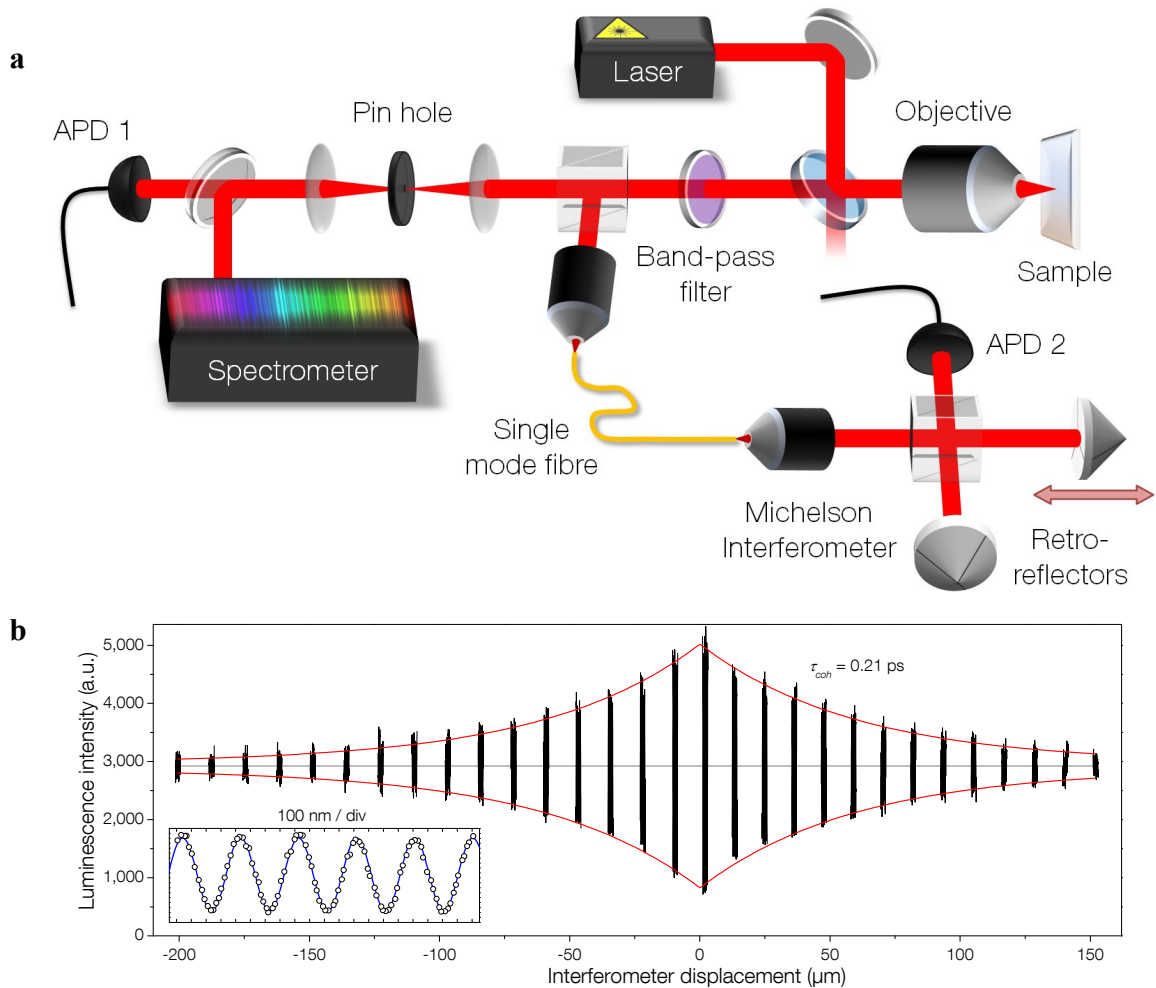

**Figure 2 | The confocal microscope and measurement of the luminescence coherence length. a**, The scanning confocal microscope collects the luminescence signal which is delivered, via a 10 metre single mode fibre, to a stabilised Michelson interferometer. The interferometer interferes a single-photon with itself to directly measure the coherence length. **b**, The resulting interference pattern envelope and single exponential fit; the inset shows a magnified view of the central interference region fitted with a sine curve.

The line-shape of the luminescence spectrum reveals a 1.6 nm full-width half-maximum Lorentzian profile indicating that photon emission was dominated by the natural lifetime and was not subject to Doppler broadening. The Lorentzian shape and known width of the emission line of the NE8 centre allowed us to make an initial estimate of the coherence properties of the emitted photons. The coherence length of the single photon state can be approximated by

$$l_{coh} = \lambda^2 / 2\pi\Delta\lambda, \qquad (1)$$

where $\Delta\lambda$ is the FWMH of the spectrum peak. Using this estimation, the coherence length implied by the luminescence spectrum was 63 μm, however factors such as phonon broadening may influence and decrease the coherence length. Until now the optical coherence properties of the NE8 centre have not be measured and speculation has been based upon measured emission linewidth.

The coherence length of the room temperature NE8 source was directly measured using a scanning Michelson interferometer (Figure 2a). By varying the path length in the interferometer about the zeroth order fringe position the measurement of the change in interference fringe contrast as a function of path length yielded a direct measurement of the single photons' coherence length. Single photons were coupled from the confocal microscope to the interferometer using a single-mode optical fibre. The automated coherence length measurements took approximately 2 hours and the sampled interference pattern from a typical scan is shown in Figure 2b. The interference pattern displays a single exponential decay of the "carrier" fringe contrast about the zeroth order fringe position. Throughout the decay envelope the fringe carrier frequency was in phase indicating the interferometer remained stable.

The shape of the fringe contrast decay curve can be better understood when one considers that the interferometer measures the Fourier transform of the source spectrum. Indeed, the Fourier transform of a Lorentzian curve is a decaying single exponential. The interferometrically measured coherence length of the NE8 emission was 63 μm (coherence time 0.21 ps). This value was derived from the $1/e$ decay point of the interference pattern envelope maxima and minima fit and is in excellent agreement with the coherence length estimated from the luminescence spectrum.

In addition to having a narrow emission linewidth one of the other advantages of the NE8 centre over other diamond colour centres is the short lifetime of the excited-state levels. Compared with the known typical lifetimes of 12-25 ns for NV and 2-3 ns for SiV we measured the NE8 centre as having a lifetime of 1.5 ns (Figure 1d). Such a short lifetime is ideal for realising high-brightness single-photon sources.

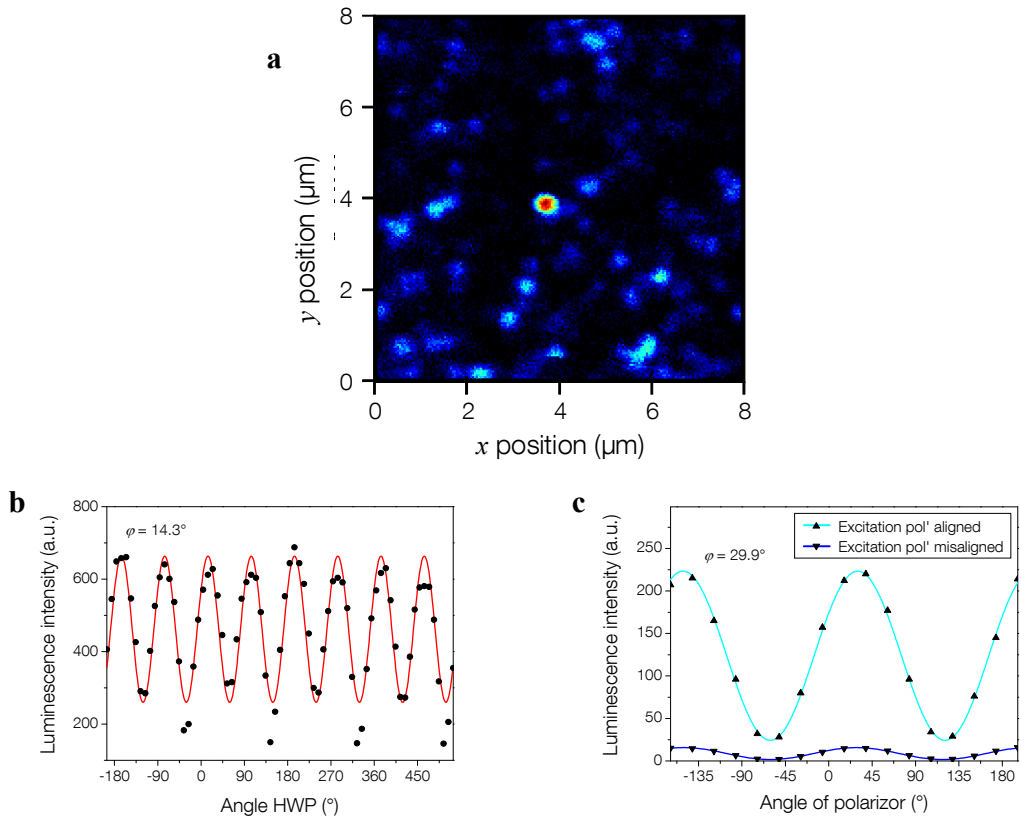

**Figure 3 | Locating NE8 centres and optical dipole characteristics. a**, A scanning confocal microscope image of the CVD-grown diamond sample and NE8 centre obtained using a band-pass filter. **b**, The NE8 luminescence signal as a function of excitation beam polarization (black points) with a cosine fit (red curve). **c**, The luminescence signal as a function of the angle of a polarizing filter for two different states of excitation polarization.

NE8 has a well defined single dipole orientation. This is in contrast to centres such as NV where the centre geometry gives rise to orthogonal transition dipoles which are incoherent [28]. It is therefore possible to maximise the NE8 centre luminescence intensity by orientating the excitation beam's polarisation with the state of polarisation of the absorption dipole. Furthermore the emission of the NE8 centre can be expected to have a single linear polarisation—a highly desirable property in future applications where the emission from separate centres is interfered. To study the polarisation properties of the absorption and emission dipole, a linear polariser was used in conjunction with a half-waveplate in the excitation beam path. Rotating the linear polarisation of the excitation beam varies the luminescence intensity sinusoidally (Figure 3b). Adding a second polarising filter in the collection arm of the microscope revealed the emission from the centre to be highly polarised and independent of the excitation polarisation (Figure 3c). The absorption and emission dipole of the centre were observed to be aligned with each other at $+29 \pm 1°$ to the vertical axis (this being the azimuthal angle, the plane of the sample representing the *xy*-plane). To measure the polar angle of the dipole we used a defocused scanned luminescence imaging technique (similar to that used to determine the orientation of emission dipoles in luminescence wide-field imaging [29]). Figure 4 shows the measured and modelled emission profile images at three defocus depths. The model confirmed the azimuthal orientation of the dipole was approximately 28° and indicated that the polar angle was 40°. This ability to measure the orientation of the NE8 dipole will be critical to future applications wherein optimised sources will require the alignment of the optical dipole with the local field of a cavity mode.

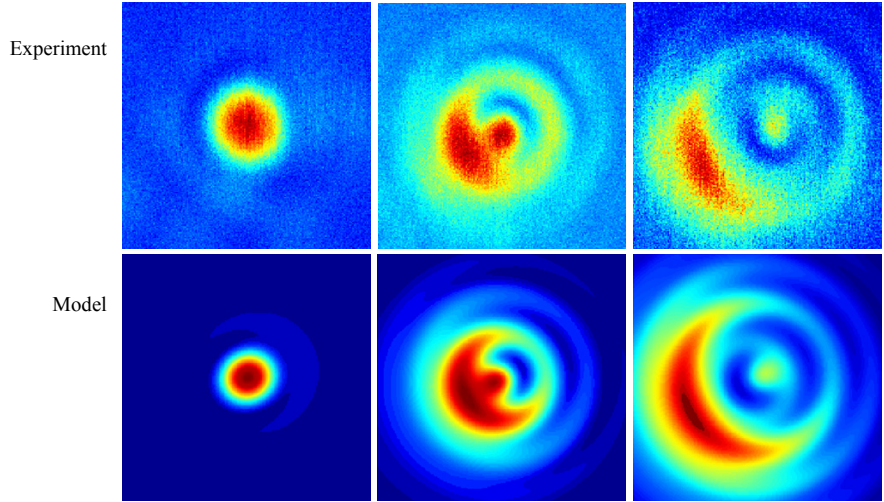

**Figure 4 | Optical dipole imaging.** Images of the dipole emission profile obtained using defocused excitation imaging (top row). The three images (from left to right) were taken at increasing defocus 'depths' of 500, 720 and 1320 nm. Theoretical model of the defocused dipole images (bottom row). The model infers a dipole orientation of $\theta = 40°$ and $\varphi = 28°$. The translation field in all images is 2 μm × 2 μm.

These results show that the orientation of the NE8 dipole can be precisely measured and that the coherence length shows promise for NE8 based transform-limited sources. The room temperature measurements show that the natural excited state lifetime $\tau_{exc}$ and coherence time $\tau_{coh}$ of NE8 are five orders of magnitude away from the time bandwidth limit given by $2\tau_{exc}/\tau_{coh} \approx 1$. Coupling to the evanescent modes of micro-spheres, toroids or photonic crystal cavities for example, will be a necessary step in developing a transform-limited room temperature source. The ability to measure dipole alignment with the electric field vector is crucial for optimizing emission efficiency of colour centres coupled to an optical cavity designed to enhance the emission rate and increase coherence length. The full potential of single colour centres in diamond for optical applications rests now on the ability to increase the spontaneous emission rate and spectrally narrow the emission line. Combining room temperature photon sources with integrated quantum circuitry [30] and the development of high efficiency single photon detectors [31], are crucial steps towards realizing scalable photonic quantum technologies.

## METHODS

The microscope used in this study (Figure 2a) consisted of a *xyz*-scanning Olympus UPlanApo 100× 1.3 NA oil immersion lens and a spatial filter containing a 50 μm pinhole. A 40 nm band-pass filter centred at 800 nm was used to filter the excitation light from the luminescence signal. A continuous-wave 685 nm laser diode was fibre coupled and its light delivered to the microscope via a 700 nm short pass filter which was used to remove any auto-luminescence originating from the coupling fibre. An uncoated quartz flat was used to reflect a portion of the excitation beam into the microscope while allowing the majority of the luminescence signal to remain in the microscope. The excitation power used during experiments was typically 500 μW (measured in front of the focusing objective).

Fabrication of the diamond samples using chemical vapour deposition has been reported previously [18]. A 3 nm bandwidth filter centred at 795 nm was used during the confocal scans for easy discrimination of the NE8 colour centres. This filter was removed for all subsequent measurements with the exception of the dipole-imaging scans.

The azimuthal angle of the dipole absorption was determined by inserting a Glan-air polarising prism in the excitation beam path in front of the turning quartz-flat. A half-wave plate (HWP) was then inserted in front of the focusing objective. The HWP imparts an angle of rotation to the linearly polarised beam that is twice the wave plate physical orientation, hence the cosine fit in Figure 3b has a phase angle, 14.3°, that is half of the absorption dipole azimuthal orientation. (The contrast of this measurement was limited by the zero-order HWP design wavelength which was 800 nm and not 685 nm.) By positioning the HWP before the quartz-flat and adjusting the laser diode current to maintain the same excitation power, the angle of the emission dipole was measured at two different linear excitation polarisations. Using a linearly polarising filter in the detector path we observed the polarisation angle of the luminescence to be independent of the angle of the excitation light's polarisation. In Figure 3c the luminescence signal intensity with polarisation angle for two different excitation polarisations is shown. The phase of the two cosine fit curves, 29.9°, is the same (within error) and is the emission dipole azimuthal orientation.

To measure the excited state lifetime ($T_1$) of the nickel colour centre we replaced the CW red excitation laser diode with a 130 ps pulsed 730 nm laser diode, the light from which, was delivered via the same optical fibre as used for the red excitation. The luminescence from the colour centre was then recorded using a time-correlated single photon counting (TCSPC) system (PicoQuant PicoHarp). After allowing for a short period when scattered light is detected by the TCSPC detector, the decay of the luminescence signal follows a single exponential curve with decay constant measured to be 1.5 ns (Figure 1c).

A Hanbury Brown and Twiss (HBT) interferometer was used to measure the second order correlation function $g^{(2)}(\tau) = \langle I(t)I(t+\tau)/I(t)^2 \rangle$ which gives an indication of the number of emitters in a particular nanocrystal. $g^{(2)}(\tau)$ is the probability of detecting two simultaneous photons (where $\tau = 0$) normalised by the probability of detecting two photons at once for a random photon source: an "antibunching" dip in $g^{(2)}(\tau)$ indicates sub-Poissonian statistics of the emitted photons and reveals the presence of a single quantum-system which cannot simultaneously emit two photons. The contrast in $g^{(2)}(\tau)$ (for this centre it was 0.65±0.03—see Figure 1c) scales as $1/N$, where $N$ is the number of emitters [11,32]. Note that in this experiment $g^{(2)}(\tau)$ was measured via the optical fibre link between the confocal microscope and interferometer APDs thereby insuring that the Michelson interferometer measured the coherence length of the single photons emitted by the NE8 centre.

To measure the coherence length of the NE8 photons the single-mode optical fibre was connected to the input of a bulk-optical Michelson interferometer comprising a 50:50 beam-splitter and two corner-cube retroreflectors (Figure 2a). To obtain a representative sample of the full NE8 interference pattern (Figure 2b) we employed a step-and-scan technique. The measurement retroreflector of the interferometer was carried on top of a motorised translation stage that included a short-range high-resolution piezo flexure. Step increments in the interferometer path length were made using the motor before scanning the retroreflector position using the piezo actuator. This technique enabled us to rapidly obtain the form of the envelope function of the interference pattern while fully resolving the interference fringes. A Michelson interferometer performs an autocorrelation of an input light field, hence the interference fringe pattern is symmetric about the zeroth order fringe and a displacement of the measurement arm mirror imparts twice the path length shift due to its retroreflective nature.

## ACKNOWLEDGEMENTS

We thank J. P. Hadden, J. Harrison, A. Politi, J. G. Rarity and M. J. Withford for their helpful discussions. Diamond samples were prepared by J.R.R. using the AsTex CVD system at the University of Melbourne. This work was supported by the Australian Research Council (through their Centres of Excellence, Discovery and Future Fellowships programs), the Australian Academy of Science, the Engineering and Physical Sciences Research Council, the Intelligence Advanced Research Projects Activity, the Leverhulme Trust, UK Quantum Information Processing Interdisciplinary Research Collaboration and the EU under IST-034368 EQUIND.

---

[*] graham.d.marshall@gmail.com
[†] james.rabeau@mq.edu.au